\documentclass[preprint,12pt]{elsarticle}

\usepackage{amssymb}

\usepackage{chemformula}
\usepackage{graphicx, float}
\usepackage{bm}
\usepackage{siunitx}
\usepackage{amsmath}

\let\oldhat\hat
\renewcommand{\vec}[1]{\boldsymbol{\mathbf{#1}}}
\renewcommand{\hat}[1]{\oldhat{\mathbf{#1}}}
\newcommand{\pd}[2]{\frac{\partial #1}{\partial #2}}
\newcommand{\zerohalf}{$\left \{ 0 0 \frac12 \right\}$ }
\newcommand{\zeroone}{$\{ 0 0 1 \}$ }
\DeclareSIUnit{\angstrom}{\textup{\AA}}  

\journal{Ultramicroscopy}

\begin{document}

\begin{frontmatter}



\title{Angstrom-Scale Imaging of Magnetization in Antiferromagnetic \ch{Fe2As} via 4D-STEM}


\author[mse]{Kayla X. Nguyen\corref{contrib}}
\author[mse]{Jeffrey Huang\corref{contrib}}
\author[mse]{Manohar H. Karigerasi}
\author[mse]{Kisung Kang}

\affiliation[mse]{organization={Department of Materials Science and Engineering, University of Illinois at Urbana-Champaign},
            city={Urbana},
            postcode={IL 61801}, 
            country={United States}}

\author[mse,mrl]{David G. Cahill}
\author[mse,mrl]{Jian-Min Zuo}

\affiliation[mrl]{organization={Materials Research Laboratory, University of Illinois at Urbana-Champaign},
            city={Urbana},
            postcode={IL 61801}, 
            country={United States}}

\author[mse,mrl,ncsa]{Andr\'{e} Schleife}

\affiliation[ncsa]{organization={National Center for Supercomputing Applications, University of Illinois at Urbana-Champaign},
            city={Urbana},
            postcode={IL 61801}, 
            country={United States}}

\author[mse,mrl]{Daniel P. Shoemaker}
\author[mse,mrl]{Pinshane Y. Huang\corref{corauthor}}
\ead{pyhuang@illinois.edu}

\cortext[contrib]{Contributed equally to this work}
\cortext[corauthor]{Corresponding author}

\begin{abstract}
We demonstrate a combination of computational tools and experimental 4D-STEM methods to image the local magnetic moment in antiferromagnetic \ch{Fe2As} with 6 angstrom spatial resolution.  Our techniques utilize magnetic diffraction peaks, common in antiferromagnetic materials, to create imaging modes that directly visualize the magnetic lattice. Using this approach, we show that center-of-mass analysis can determine the local magnetization component in the plane perpendicular to the path of the electron beam.  Moreover, we develop Magnstem, a quantum mechanical electron scattering simulation code, to model electron scattering of an angstrom-scale probe from magnetic materials. Using these tools, we identify optimal experimental conditions for separating weak magnetic signals from the much stronger interactions of an angstrom-scale probe with electrostatic potentials.  Our techniques should be useful for characterizing the local magnetic order in systems such in thin films, interfaces, and domain boundaries of antiferromagnetic materials, which are difficult to probe with existing methods.
\end{abstract}




\end{frontmatter}


\section{Introduction}
\label{sec:intro}
For decades, transmission electron microscopy (TEM) and scanning TEM (STEM) methods such as Lorentz TEM, electron holography, and differential phase contrast (DPC) have provided valuable insight into the magnetic structure of materials on the micro and nanoscales. The relatively high spatial resolution of these methods, which is typically 5-20 nanometers (nm), but can reach up to around 1 nm \cite{Boureau2021}, have made them valuable for uncovering variations in the local magnetization such as domain walls in ferromagnets, skyrmions, and the magnetization of nanoparticles and thin films. As TEM-based techniques continue to improve, a natural question is whether electron microscopy techniques can be extended to probe local magnetization on the atomic scale.

\begin{figure}[bp!]
\includegraphics[width=0.95\textwidth]{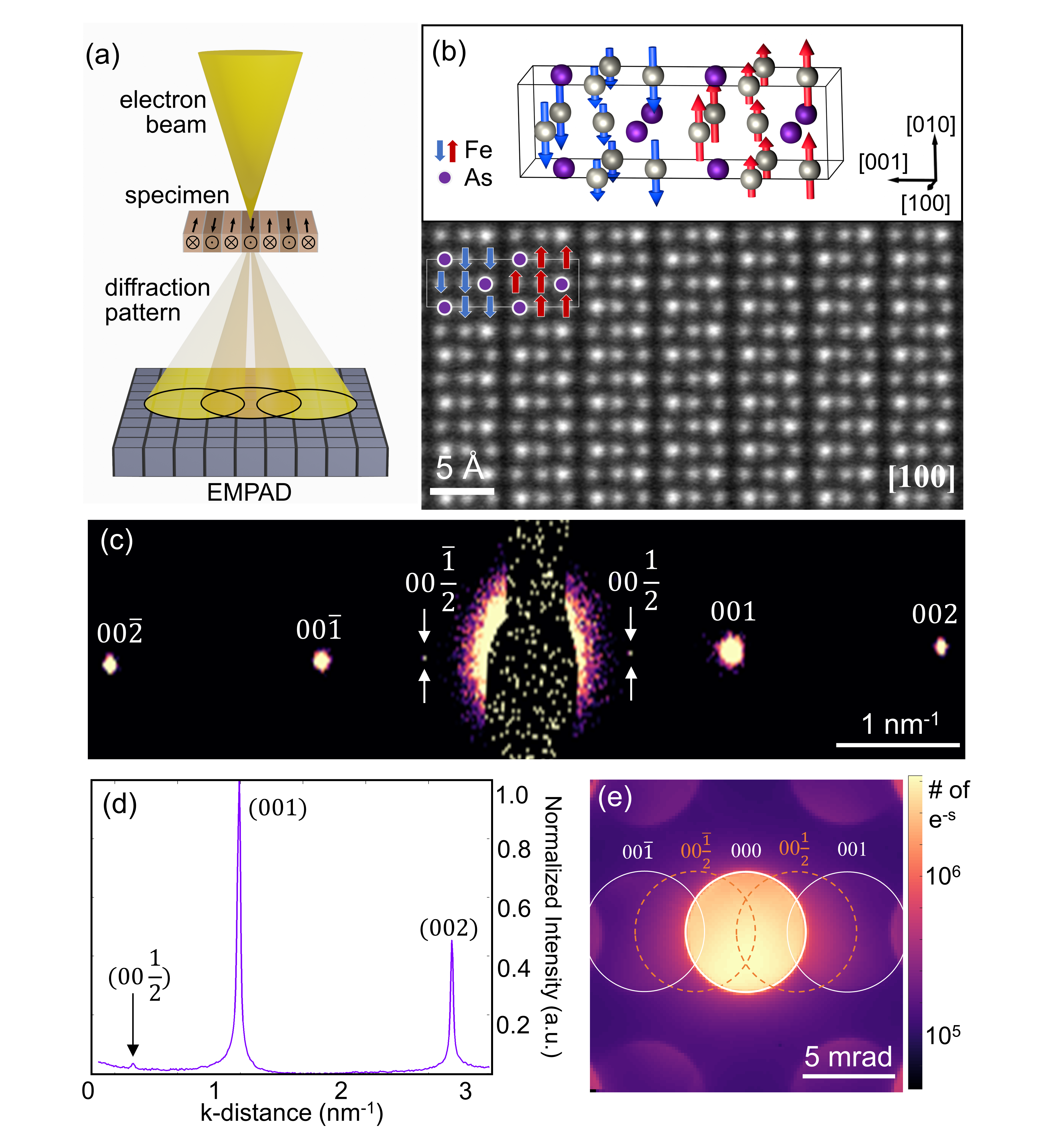}
\caption{\label{fig:ExpSetup} Detecting antiferromagnetism in \ch{Fe2As} using electron diffraction. (a) 4D-STEM schematic showing diffraction from an antiferromagnet with an angstrom-size probe. (b) Magnetic unit cell (top) and atomic resolution ADF-STEM image (bottom) of \ch{Fe2As}. Arrows indicate directions of Fe magnetic moments. (c) TEM diffraction pattern showing weak magnetic \zerohalf spots. The complete diffraction pattern is shown in Figure~S1. (d) Line profile of (c), showing the relative intensity of each peak. (e) CBED pattern captured in 4D-STEM with circles indicating diffracted disks.}
\end{figure}

Antiferromagnets such as \ch{Fe2As} (Figure~\ref{fig:ExpSetup}b) have received renewed interest as potential materials to increase the speed and density of spintronic devices because they exhibit terahertz spin dynamics, have low magnetic susceptibility, and lack stray fields. Because their local magnetization varies periodically on the scale of a few angstroms, antiferromagnets offer ideal systems to test methods for atomic scale magnetic characterization. New TEM-based magnetic imaging techniques could directly probe magnetic ordering in antiferromagnets in real space,  complementing bulk characterization methods such as neutron diffraction \cite{SchlenkerXray}, while also achieving the high spatial resolution of surface sensitive techniques such as spin-polarized scanning tunneling microscopy\cite{Enayat653}.

Yet, few TEM studies have been attempted to image antiferromagnetic order. Part of the challenge is spatial resolution: a few-nanometer electron beam would average over several magnetic unit cells in most antiferromagnets, producing no net magnetic deflection. Sensitivity is another: for techniques such as DPC that detect magnetization as a deflection of the electron beam, the minimum detectable deflection is inversely proportional to probe size\cite{CHAPMAN1978203, Nguyen_PRApplied}. This trade-off between spatial resolution and sensitivity to magnetic fields makes it difficult to detect magnetic signals at atomic length scales. A third and particularly insidious challenge for electron microscopy is that magnetization in antiferromagnets varies on the same length scale as the atomic spacings in materials, making it exceptionally difficult to isolate magnetic signals from scattering effects from the crystallographic lattice. 
Recently, DPC-STEM has been used to probe antiferromagnetism across a domain wall \cite{AFM_ScienceAdvances2022} and, via kernel filtering and unit-cell averaging, has recently been used to map the magnetic structure of an antiferromagnet in real space \cite{Kohno2022}.  However, these methods remain somewhat indirect, and the detection mechanisms behind these recent observations and how magnetic signals can be reliably separated from atomic structure remain unclear.   By collecting the full diffraction pattern, four-dimensional (4D) STEM studies can provide greater measurement flexibility and insight into the effect of magnetism on electron scattering.  Here, we use the latest generation of direct electron detectors for 4D-STEM, combined with electron scattering simulations to model, measure, and extract the local magnetization on the few-angstrom scale.

\section{Results and Discussion}
We demonstrate our approach on \ch{Fe2As}, a collinear, metallic antiferromagnet with a N\'eel temperature of \SI{353}{\kelvin}. Figure 1b shows the structure of \ch{Fe2As}: it has a tetragonal crystal structure ($a= \SI{3.63}{\angstrom}$, $c = \SI{5.98}{\angstrom}$) \cite{Katsuraki_doi:10.1143/JPSJ.21.2238}.  The N\'eel vector for \ch{Fe2As} can lie along either of the crystallographically equivalent $a$ or $b$ axes, and the magnetic unit cell has dimensions $a \times a \times 2c$. Single crystals of Fe$_2$As were grown by slow-cooling stoichiometric mixtures of the elements from the melt at 1000\,°C in fused silica ampoules, following the detailed procedure in Ref.\ \citenum{ShoemakerFe2As}. Then, TEM specimens of \ch{Fe2As} were prepared using standard focused ion beam lift-out procedures and imaged with aberration-corrected STEM.  Figure 1b shows an annular dark-field STEM image of \ch{Fe2As} along the [100] zone axis. In this orientation, vertical blocks three atoms wide, or ``trilayers'' are visible.  Within each \SI{6}{\angstrom}-wide trilayer, the magnetic moments of the Fe atoms are aligned parallel, while the magnetic moments of adjacent trilayers are antiparallel.

\subsection{Selected area electron diffraction}
We first confirm that antiferromagnetic ordering can be detected in our sample. Figure 1c shows a room temperature selected area electron diffraction (SAED) pattern of \ch{Fe2As} along the [100] zone (see full diffraction pattern in Figure~S1).   While the brightest peaks in the diffraction pattern arise from Bragg diffraction from the crystal, we also observe weak diffraction spots (white arrows in Figure 1c) with an intensity of roughly 1/20 to 1/300 of the \zeroone diffraction spots (Figure 1d), depending on the region and sample thickness measured. These magnetic \zerohalf diffraction peaks correspond to the $2c$ length of the magnetic unit cell. Similar half-order peaks have been previously observed in electron beam diffraction of antiferromagnetic \ch{NiO} \cite{LoudonPRL}, and magnetic diffraction peaks with fractional k-vectors are also routinely observed in neutron diffraction studies of antiferromagnets, where they are used to determine magnetic structures. Importantly, the presence of the \zerohalf spots in the SAED pattern demonstrates that antiferromagnetic ordering is detectable in our \ch{Fe2As} sample.

\subsection{Experimental 4D-STEM}
Next, we extend our measurements to the atomic scale using 4D-STEM.  In 4D-STEM, an angstrom to nanoscale electron beam is rastered across a sample, and a convergent beam electron diffraction (CBED) pattern is collected at each scan position (Figure 1a). When combined with a new generation of fast, high-dynamic range direct electron detectors, 4D-STEM offers new opportunities to probe the structure and functional properties of materials at the atomic scale. A key advantage of 4D-STEM is that it collects the two-dimensional (2D) electron scattering distribution as a function of position, offering atomic-scale resolution while also providing more flexibility to extract and isolate signals of interest. This property has enabled measurements of angstrom-scale electrostatic fields, such as the local polarization of ferroelectrics \cite{Yadav2019,ferroelectric_reviewdoi:10.1063/5.0035958,ophus_2019} as well as nanoscale mapping of magnetic fields\cite{Nguyen_PRApplied}. 

To acquire our 4D-STEM data, we used a Thermo Fisher Titan equipped with an electron microscope pixel array detector (EMPAD), a 128$\times$128 pixel direct electron detector with a high dynamic range of 1,000,000:1\cite{tate_EMPAD}.  This high dynamic range allows us to achieve high signal-to-noise ratios and detect small intensity changes from magnetism\cite{Nguyen_PRApplied}.  We used an acceleration voltage of 120 kilovolts (kV) and a semi-convergence angle of 3.5 milliradians (mrad), yielding a probe size of approximately $\SI{6}{\angstrom}$. This probe size was chosen to approximately match the width of the trilayers and the length scale of magnetization variations in our specimen. As we discuss later, this convergence angle maximizes the magnetic signal.
We also operated with the objective lens on, rather than in a Lorentz or field-free mode.  In this mode, the objective lens produces a magnetic field of 1.61 T normal to the sample plane, which should orient the Fe$_2$As moments in-plane, perpendicular to the beam direction\cite{Kexin_PhysRevB}.  We calibrated the strength of the objective lens using Hall probe measurements reported in Ref.~\citenum{Nguyen_PRApplied}.
Figure 1e shows a position averaged CBED pattern of \ch{Fe2As} along the [100] zone axis captured using 4D-STEM. The CBED pattern contains similar information to the SAED pattern in Figure 1c, but the use of a convergent beam results in the formation of diffraction disks, rather than spots. As we show below, while the magnetic \zerohalf reflections are too weak to be directly visible in the CBED pattern (Figure 1e), they nevertheless produce measurable signals that enable atomic scale magnetic imaging.


\begin{figure}[H]
\includegraphics[width=0.95\textwidth]{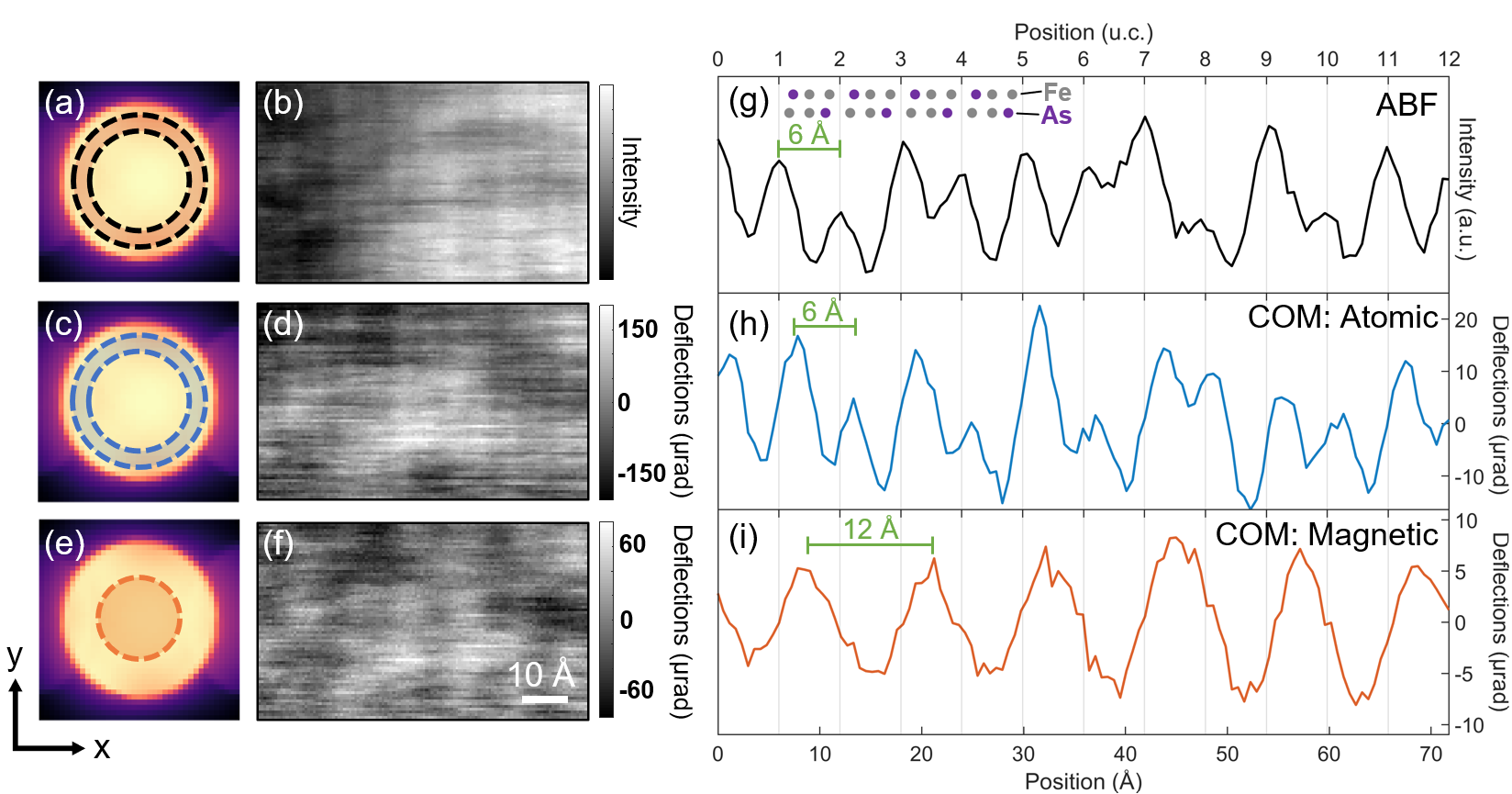}
\caption{\label{fig:ExpResults} Separating structural and magnetic lattices via COM imaging. Virtual ABF and COM images from experimental 4D-STEM data, acquired using an electron beam with 3.5 mrad semi-convergence angle at 120 kV, over 120 x 71 probe positions with a spacing of \SI{0.61}{\angstrom} per pixel.  Diffraction patterns and virtual images are oriented with the $c$ axis pointing in the x-direction and an $a$ axis pointing in the y-direction. (a) An annular virtual detector is used to reconstruct (b) an annular bright field (ABF) image. Integrating over (c) the same annular region as (b), (d) an x-component center of mass (COMx) image showing atomic contrast is formed.  Integrating over (e) a small circular region within the BF disk to form (f) a COMx image showing magnetic contrast. (g-i) Background-subtracted line profiles of the images (b), (d), and (f). The overlay in (g) shows the atomic structure of \ch{Fe2As}. COMx deflections due to magnetic contrast have a spatial period twice that of the atomic contrast, consistent with the structure of \ch{Fe2As}. Line profiles were obtained by averaging along the (001) lattice fringes seen in Fig. 2(b) to mitigate the effects of scan distortion.}
\end{figure}

In Figure 2, we extracted three signals of interest from a 4D-STEM dataset.  First, we created a virtual annular bright field (ABF) image by applying a digital mask (Figure 2a) to each CBED pattern and integrating the counts within the virtual detector at each scan position.  The resulting image (Figure 2b) is similar to that generated by a conventional physical ABF STEM detector; dark vertical lines correspond to the locations of trilayers. Compared to the ADF STEM image in Figure 1b, the contrast in the virtual ABF image is inverted because of the different choice of collection angles and the image resolution is decreased due to the relatively large \SI{6}{\angstrom} probe size.

In Figure 2c-f, we apply center-of-mass (COM) analysis to our 4D-STEM data to separate and isolate signals from the crystal lattice and magnetic structure.  COM analysis measures the center of mass of the intensity of the diffraction patterns acquired via 4D-STEM.  When applied to the full diffraction pattern, the COM measures changes to the average momentum of the electron beam produced by its interactions with the specimen; correspondingly, COM analysis is commonly used to measure electric fields and magnetic domains \cite{MullerK2014,mcao_theory_empad_2018,Nguyen_PRApplied}.  In addition, application of COM analysis to selected regions of the CBED pattern has been shown to be an effective method to separate scattering phenomena that dominate different regions of reciprocal space \cite{Nguyen_PRApplied}.

While the COM is a 2D vector quantity, we focus here on a single component, COM shifts along the $c$-axis direction, since we expect that both the electrostatic and magnetic fields of \ch{Fe2As} will produce COM deflections along this direction. Orienting the real space images and diffraction pattern so that the $c$ axis points in the x-direction, we calculate the $\mathrm{COM}_x$ shifts in regions of the diffraction pattern defined by the masks shown in Figure 2c,e, creating the COM images in Figure 2d,f.  Figure 2g-i show background subtracted line profiles of the virtual ABF (Figure 2g) and two COM images (Figure 2h,i). 
Both the ABF line profile in Figure 2g and the COM data in Figure 2h reveal lattice fringes with the \SI{6}{\angstrom} periodicity of the trilayers of the structural unit cell.  The COM signal in Figure 2i is especially interesting: it alternates in direction from one trilayer to the next, and exhibits a \SI{12}{\angstrom} periodicity. As we show in Figure S3, these fringes can also be detected using DPC imaging using a virtual quadrant detector. The position and frequency of this COM signal are an excellent match for the local magnetization changes expected for antiferromagnetic \ch{Fe2As}, a strong indicator that our data represents an experimental detection of the atomic scale antiferromagnetic ordering. 

\begin{figure}[H]
\includegraphics[width=0.95\textwidth]{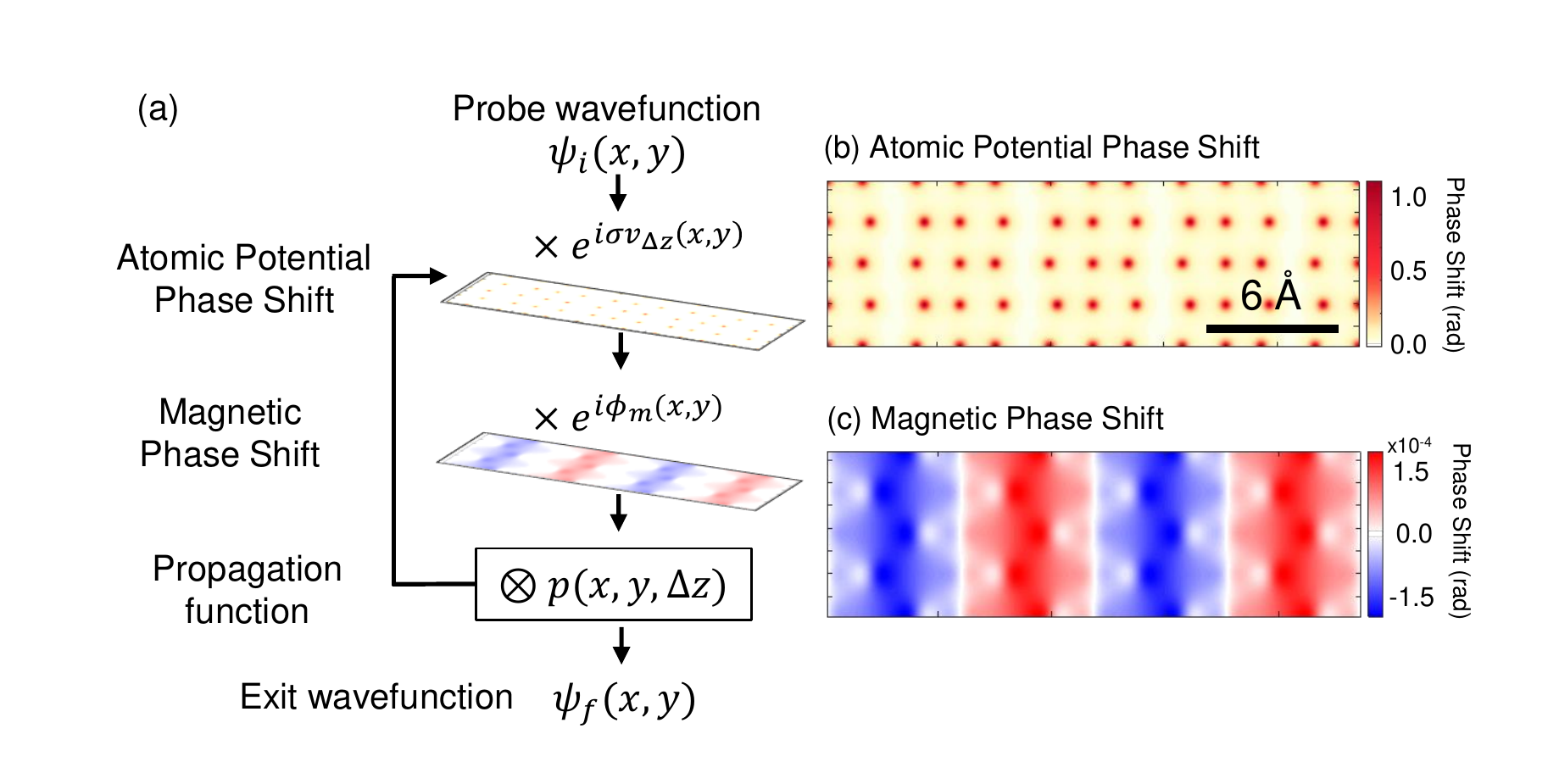}
\caption{\label{fig:SimSetup} Schematic of the Magnstem approach for electron scattering simulations in magnetic specimens: (a) The multislice method is modified to use a transmission function consisting of two phase shifts to model scattering through each atomic layer: a phase shift due to atomic potentials and a magnetic phase shift.  (b) Phase shift due to projected atomic potentials and (c) phase shift due to projected magnetic fields shown for a one-unit-cell thick region of [100] \ch{Fe2As}. As the probe wavefunction travels through the simulated sample, it receives a phase shift and is convolved with the propagation function at each slice of the sample.}
\end{figure}

\subsection{4D-STEM simulations using Magnstem}
To understand the origins of these signals, we developed Magnstem, a quantitative simulation code to model the electron scattering of an angstrom-scale probe in magnetic materials. We based our methods on multislice \cite{CowleyMoodie1957}, a powerful approach for simulating TEM experiments and calculating the dynamic scattering of the electron beam from electrostatic fields in the specimen. In multislice, the sample is divided into thin slices and the electron wave function is alternately scattered by the transmission function of a slice and propagated towards the next slice, as shown in Figure 3a. While multislice simulations typically focus only on scattering from the screened nuclear potential, a few groups have adapted multislice codes to incorporate scattering from magnetic fields \cite{Rother2009,Grillo2013,Edstrom2016Vortex,Rusz_QuantumDPC}.

In Magnstem, we describe the evolution of the probe wavefunction as follows:
\begin{equation}
\pd{\psi(\vec{r})}{z} = \left[ \frac{i}{2k} \nabla^2_{xy} + i \left( \frac{me}{\hbar^2k} V(\vec{r}) - \frac{e}{\hbar}A_z(\vec{r}) \right) \right] \psi(\vec{r}).
\label{eq:sch}
\end{equation}  Here, the effect of magnetic fields within the plane of the sample enters through the component of the vector potential along the propagation direction $A_z$. We retain only the $A_z$ term from the full paraxial Pauli equation for a magnetic sample (Eq.\ (3) in Ref.\ \citenum{Edstrom2016Vortex}) because this term is two orders of magnitude larger than other terms that describe the interaction between electrons and magnetic fields when the spins are oriented perpendicular to the electron beam (see the Supplementary Material). Equation \ref{eq:sch} can then be modeled using the conventional multislice method using the following transmission function for a slice of the sample,
\begin{equation}
    t(x, y, z) = \exp \left[ i \int_{z}^{z+\Delta z} \left( \sigma V(x, y, z') - \frac{e}{\hbar}A_z(x, y, z') \right) dz' \right]
\end{equation}
where $\sigma = me/\hbar^2 k$.

To implement this method, we modified the Kirkland multislice code \cite{Kirkland2010}, adding a contribution to the transmission function due to $A_z$ (shown in Figure 3c), which we derived from magnetization densities calculated for \ch{Fe2As} using density-functional theory \cite{ShoemakerFe2As} (see the Supplementary Material). Magnstem adds to conventional multislice, where the electron beam undergoes phase shifts due to scattering from atomic potentials (Figure 3b), by incorporating an additional phase shift due to the in-plane magnetic moments of the sample (Figure 3c). As Figure 3b-c show, the additional phase shift due to magnetic moments is small: 1,000 to 10,000 times smaller than the phase shift due to atomic moments.

We next use Magnstem to simulate 4D-STEM datasets, matching key parameters of simulation to experiment (see the Supplementary Material).  We simulate \ch{Fe2As} down the [100] direction with a thickness of 9.1 nm (25 unit cells), creating a 4D dataset by rastering the simulated probe over $15 \times 45$ probe positions, saving the CBED pattern at each probe position.
To account for atomic vibrations, the frozen phonon approximation was employed \cite{Kirkland2010phonons}, with each CBED pattern averaged over 10 frozen phonon configurations.
We note that this approach is distinct from the finite difference technique to compute phonon properties of materials from first-principles \cite{Kunc1985}, sometimes also referred to as frozen-phonon method.
To isolate the impact of antiferromagnetism on electron scattering, we conducted two sets of 4D-STEM simulations with the magnetism turned either on or off, then used principal component analysis (PCA) to analyze the resulting CBED patterns (Figure 4).

\begin{figure}[H]
\includegraphics[width=0.95\textwidth]{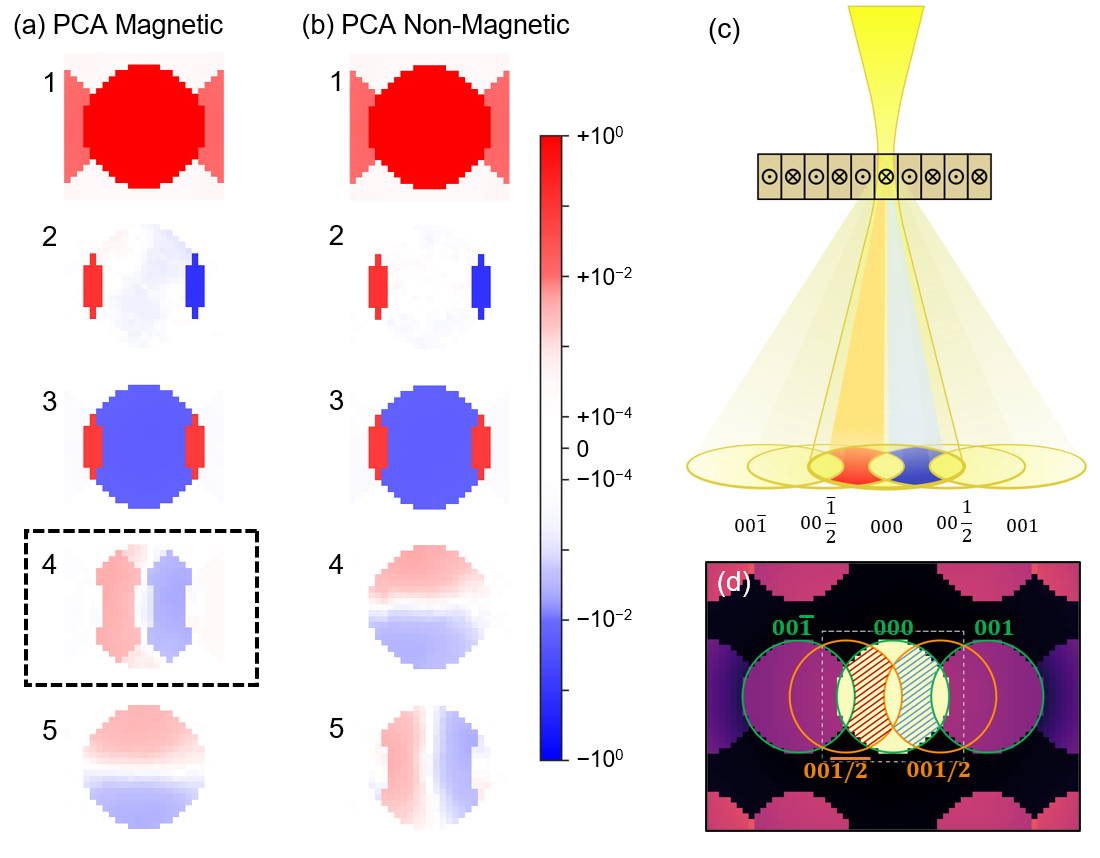}
\caption{\label{fig:PCA} Principal component analysis of electron scattering simulations shows unique principal component from magnetism. (a, b) The five largest principal components of the CBED patterns of a simulated 4D-STEM data set with magnetism  (a) turned on or (b) turned off.  The principal components are scaled by their singular values and arranged in descending order by size. The fourth component from the magnetic simulation results from the magnetic lattice, and its symmetry corresponds to the overlap between the direct beam and \zerohalf disks. (c) Schematic of the impact of the magnetic PCA component on the 4D-STEM signal. (d) Simulated PACBED pattern where the \zeroone disks are outlined in green and \zerohalf disks are outlined in orange.  The expected asymmetry from magnetism can be observed in the BF disk within the overlap region of the \zerohalf disks (red and blue hatched areas).  The region of the CBED patterns used in PCA analysis is indicated by the white dashed rectangle.}
\end{figure}

Figures 4a-b show the principal components for the two 4D-STEM data sets.  Comparing the magnetic and non-magnetic simulations, the first three principal components and their corresponding real-space maps are nearly identical, indicating they arise primarily from the atomic lattice of the material (see Figure~S4 for PCA components, singular values, and corresponding real space maps).  Accordingly, real space maps of the first three PCA components (Figure~S4) each produce (001) lattice fringes with the \SI{6}{\angstrom} periodicity of the crystallographic unit cell.

Notably, we find that the fourth PCA component of the magnetic simulations (Figure 4a, black dashed box) provides the basis for angstrom-scale STEM imaging of antiferromagnetism.  The shape and symmetry of this component, which is absent in the non-magnetic simulations, corresponds to regions of overlap between the BF disk and the magnetic \zerohalf disks (Figure 4c-d).  In addition, real space maps for this component (Figure S4) show $\left( 0 0 \frac12 \right)$ lattice fringes with the \SI{12}{\angstrom} periodicity of the magnetic unit cell.

These results are important because they explain the physical origin behind angstrom-scale detection of antiferromagnetism and illustrate how to set up direct imaging modes for magnetic lattice fringes.  In short, the periodic magnetic structure of \ch{Fe2As} acts as a weak phase grating, producing \zerohalf disks. When these magnetic reflections overlap with the direct beam, they produce  magnetic lattice fringes--a direct parallel to how overlap between structural Bragg disks gives rise to structural lattice fringes.  Notably, this mechanism is distinct from the conventional mechanisms used to explain Lorentz TEM and STEM signals for ferromagnetic materials, where slowly varying magnetic fields produce rigid shifts of the entire diffraction pattern. The semi-convergence angle has a strong impact on the interpretability and strength of the magnetic signal. We show using Magnstem simulations (Figure~S5) that the magnetic signals are strongest when the probe size is approximately \SI{6}{\angstrom} wide, matching the length scale of the magnetic columns in the sample. This can also be understood in terms of the diffraction pattern: magnetic contrast arises from overlap of the magnetic disks with the BF disk, and this probe size maximizes that overlap (Figure~S6).

\begin{figure}[bp!]
\includegraphics[width=0.65\textwidth]{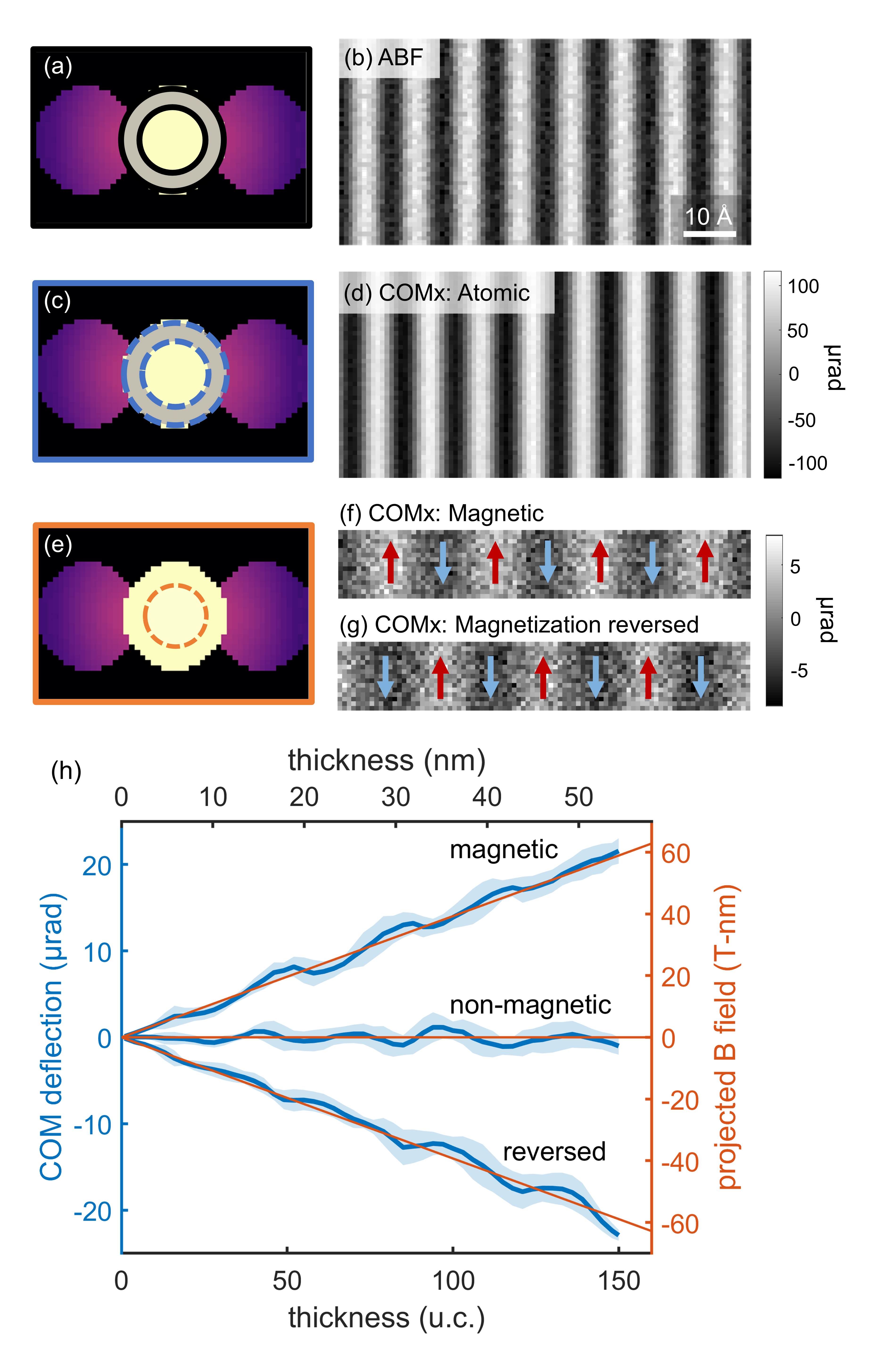}
\caption{\label{fig:SimCOM} Center-of-mass imaging in simulated 4D-STEM data. (a) An annular virtual detector, shown overlaid over a simulated CBED pattern, is used to reconstruct (b) an ABF image. (c-g) Two images mapping the x-direction shifts in the center of mass (COMx) were constructed, using either (c-d) an annular region which overlaps with the \zeroone disks or (e-g) a small circular region that overlaps with the \zerohalf disks. (f-g) In (g), the magnetic moments were reversed in direction compared to (f).  The contrast in (g) is opposite that of (f), indicating that the direction of the in-plane magnetic moment can be uniquely determined. The ABF and COMx images were tiled to emphasize the spatial periodicity of the fringes. (h) COMx deflection (blue) as a function of sample thickness, for simulated thicknesses up to 150 unit cells. COMx deflection values were averaged over 8 probe positions along the middle of a trilayer; the standard deviation from those probe positions is shown as blue shading.}
\end{figure}


Finally, we performed COM analyses of the 4D-STEM simulations, mirroring the processing approach used for experimental data. Unlike in simulations, we found that a simple PCA analysis was not sufficient to isolate the magnetic signal in experimental data. This is likely because CBED patterns from experimental samples contain spatially-varying components due to sample preparation and beam damage making it difficult to isolate the weaker magnetic contributions in the CBED patterns. Overall, our simulations are an excellent match for the experimental data in Figure 2. In contrast to PCA, we found COM imaging using selected regions of the BF disk to be a robust approach for isolating magnetism in both experiment and simulations. Moreover, we show in Figure 5f-g that COM analysis can unambiguously determine the sign of the in-plane component of magnetic moment (for example, the up vs.\ down arrows in Figure 5f).  This ability arises from the asymmetry of the phase between the \zerohalf peaks and the corresponding intensity asymmetry of the magnetic PCA component in Figure 4. 



In TEM imaging of crystalline samples, lattice fringes arising from the atomic structure have been observed to reverse in contrast as the sample thickness changes. Thus, proper interpretation of the magnetic signal requires understanding how it scales with sample thickness. We simulated 4D-STEM data sets over a range of thicknesses, from 0.36 nm to 54.3 nm (Figure 5h) and measured the intensity of magnetism-induced COM shifts in the overlap region (Figure 5e). Figure 5h shows that the COMx signal increases approximately linearly with the sample thickness and projected magnetic field. These results show that it is advantageous to use a thicker specimen to maximize the magnetic signal, without potential complications from contrast reversal.

\section{Conclusions}
In summary, we developed an approach for imaging the magnetic structure of antiferromagnets in real space using 4D-STEM. In our work, to increase the signal from the magnetic structure, we match the size of the electron probe to the size of regions of aligned spins. Using Magnstem simulations, we observe that the periodic magnetic structure of \ch{Fe2As} redistributes intensity in CBED patterns at the overlap of antiferromagnetic \zerohalf disks with the BF disk. By selecting those regions within the BF disk, COM imaging can be used to detect the magnetic structure of \ch{Fe2As}. This approach can be broadly applied to antiferromagnets whose magnetic structures produce fractional antiferromagnetic reflections, and should allow for the determination of magnetic structure in antiferromagnetic thin films and imaging of antiferromagnetic domains, antiphase boundaries, and domain walls.

\section*{Declaration of Competing Interest}
KXN is a co-inventor on the license for the EMPAD detector to Thermo Fisher Scientific.

\section*{Acknowledgements}
The authors thank D. A. Muller for useful discussions. This work was supported by the Air Force Office of Scientific Research under award number FA9550-20-1-0302. K.X.N. is also supported in part by the L'Oreal For Women in Science Postdoctoral Fellowship. J.H. is partially supported by the DIGI-MAT program through National Science Foundation under Grant No. 1922758. Electron microscopy facilities were provided by the University of Illinois Materials Research Laboratory and the Cornell Center for Materials Research through the National Science Foundation MRSEC program through award DMR-1719875.  The crystal growth and density functional calculations were supported by the Illinois Materials Research Science and Engineering Center through the National Science Foundation MRSEC program, award DMR-1720633.
This work made use of the Illinois Campus Cluster, a computing resource that is operated by the Illinois Campus Cluster Program (ICCP) in conjunction with the National Center for Supercomputing Applications (NCSA) and which is supported by funds from the University of Illinois at Urbana-Champaign.

\section*{Supplementary materials}
Supplementary material, including Material and Methods and additional figures, are appended to the end of this manuscript file.




 \bibliographystyle{elsarticle-num} 
 \bibliography{fe2aspaper}





\end{document}


\begin{frontmatter}



\title{Supplementary Material: Angstrom-Scale Imaging of Magnetization in Antiferromagnetic \ch{Fe2As} via 4D-STEM}


\author[mse]{Kayla X. Nguyen\corref{contrib}}
\author[mse]{Jeffrey Huang\corref{contrib}}
\author[mse]{Manohar H. Karigerasi}
\author[mse]{Kisung Kang}

\affiliation[mse]{organization={Department of Materials Science and Engineering, University of Illinois at Urbana-Champaign},
            city={Urbana},
            postcode={IL 61801}, 
            country={United States}}

\author[mse,mrl]{David G. Cahill}
\author[mse,mrl]{Jian-Min Zuo}

\affiliation[mrl]{organization={Materials Research Laboratory, University of Illinois at Urbana-Champaign},
            city={Urbana},
            postcode={IL 61801}, 
            country={United States}}

\author[mse,mrl,ncsa]{Andr\'{e} Schleife}

\affiliation[ncsa]{organization={National Center for Supercomputing Applications, University of Illinois at Urbana-Champaign},
            city={Urbana},
            postcode={IL 61801}, 
            country={United States}}

\author[mse,mrl]{Daniel P. Shoemaker}
\author[mse,mrl]{Pinshane Y. Huang\corref{corauthor}}
\ead{pyhuang@illinois.edu}

\cortext[contrib]{Contributed equally to this work}
\cortext[corauthor]{Corresponding author}

\end{frontmatter}
\tableofcontents

\renewcommand\thepage{S-\arabic{page}}
\renewcommand{\thefigure}{S\arabic{figure}}

\setcounter{figure}{0}
\section{Selected area diffraction pattern}
\begin{figure}
    \includegraphics[width=.85\textwidth]{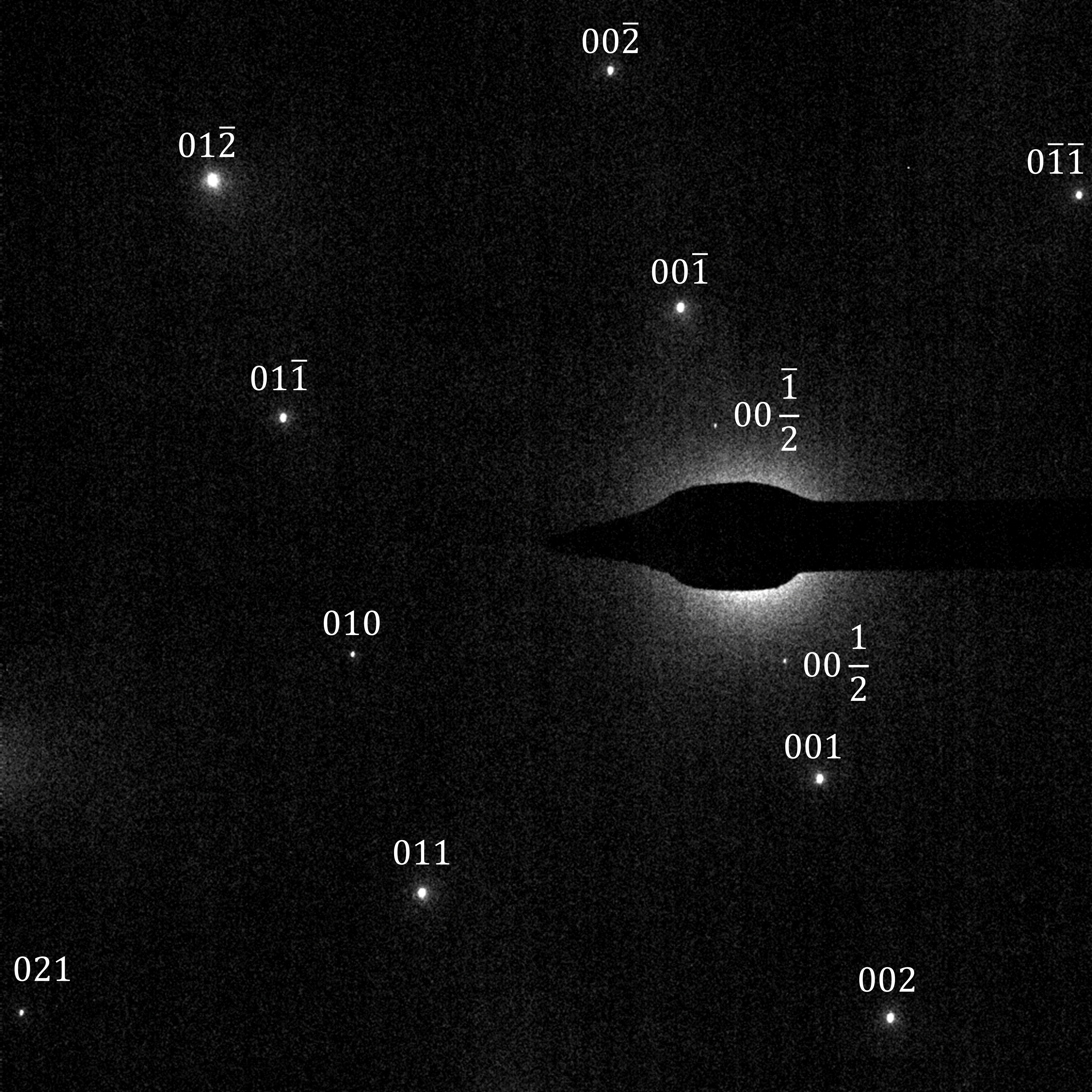}
    \caption{Selected area electron diffraction pattern of \ch{Fe2As}.}
    \label{fig:SAED}
\end{figure}

Selected area electron diffraction (SAED) pattern (Figure S1) acquired using a Thermo Fisher Themis Z at 300 kV with a Ceta CMOS camera.

\section{Experimental 4D-STEM data}
The 4D-STEM data set shown in Figure 2 was cropped from a larger data set of 256 x 256 probe positions.  The thickness of the \ch{Fe2As} sample is estimated to be about 30-45 nm by comparing the PACBED pattern against multislice simulations.

To construct the line profiles in Figure 2g-i, first, a high quality image of (001) lattice fringes was reconstructed from the larger data set by forming a x-component center of mass (COMx) image that integrated over regions of overlap between the bright field disk and the $\{ 001 \}$ disks. The lattice fringes are not perfectly straight, indicating the presence of scan distortion or sample drift. This image was then Fourier filtered to select for the (001) lattice fringes.  Then, the phase of the (001) lattice fringes was unwrapped; the unwrapped phase serves as an estimate of position along the c-axis of the \ch{Fe2As} sample.  Finally, pixels within the selected region of interest were binned according to unwrapped phase, and averaged within each bin to produce the line profiles.

\begin{figure}
    \includegraphics[width=.75\textwidth]{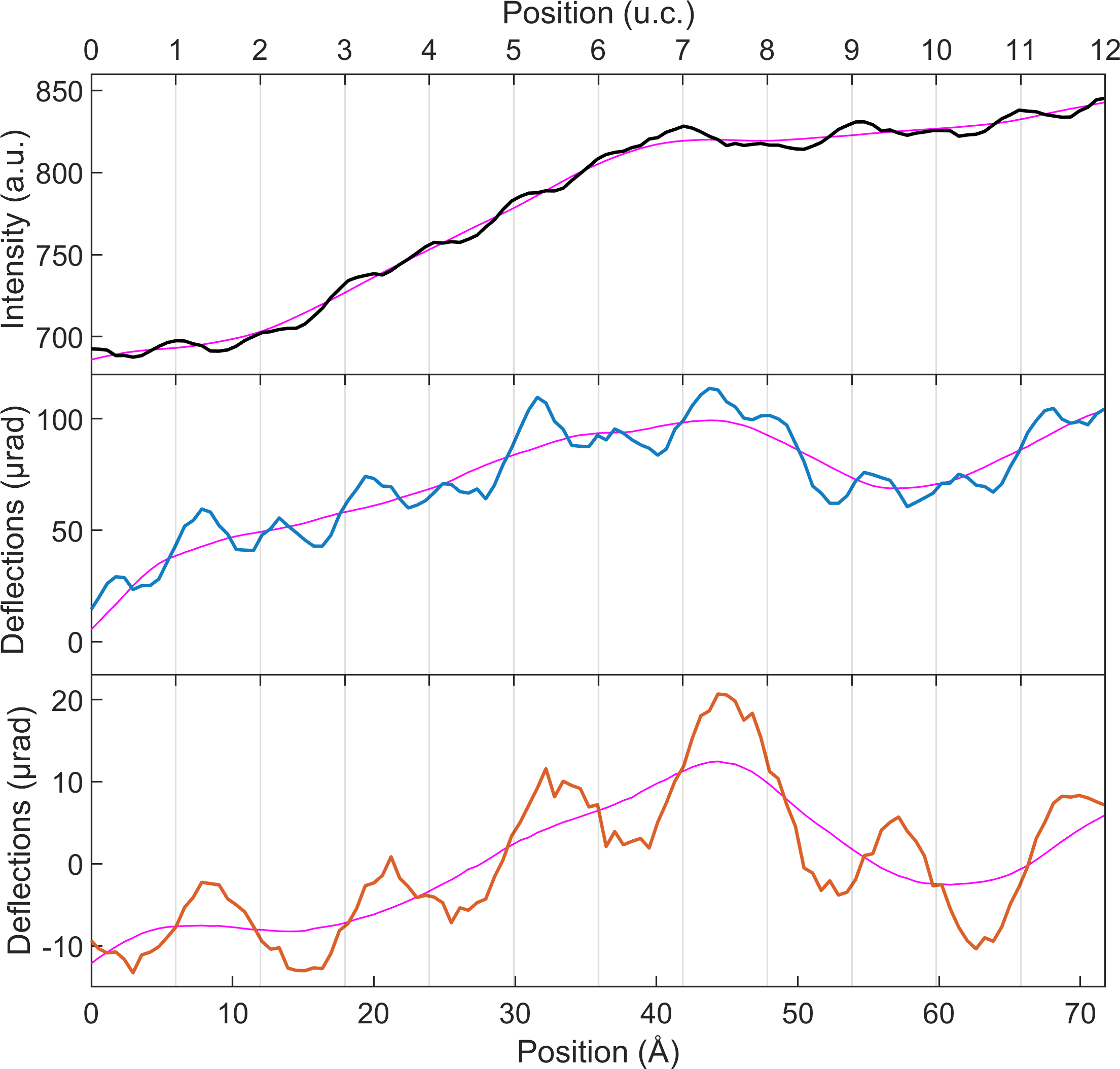}
    \caption{Background subtraction of line profiles of the real space images in Figure 2.  For each line profile, a slowly varying background (magenta curves) is computed as a moving average.}
    \label{fig:LineProfSupp}
\end{figure}

Furthermore, the line profiles were background subtracted. A background is computed for each line profile as a centered moving average with a window length of two unit cells.  This background is subtracted from the line profile.  The backgrounds and the line profiles before background subtraction are shown in Figure \ref{fig:LineProfSupp}.

\subsection{Virtual differential phase contrast imaging}
\begin{figure}[h]
    \includegraphics[width=.95\textwidth]{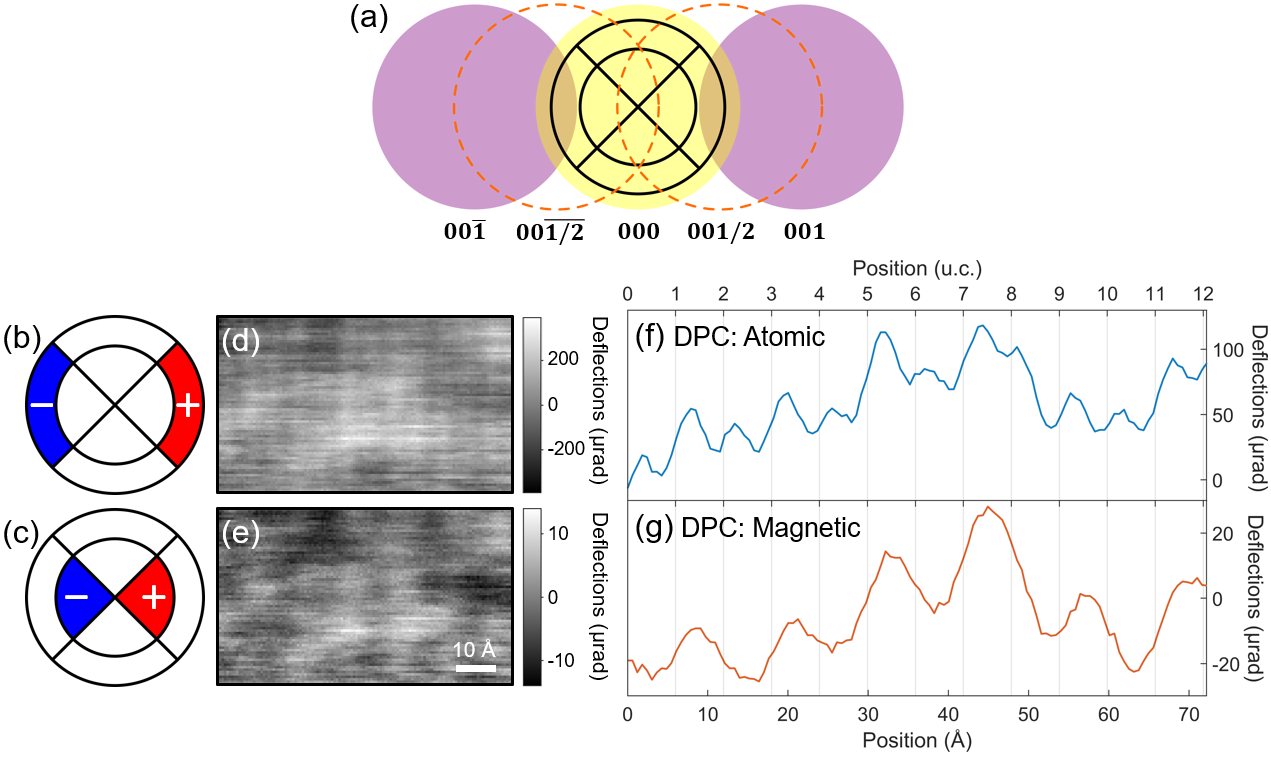}
    \caption{\label{fig:DPC}Virtual Differential Phase Contrast (DPC) Analysis. (a) Schematic showing position of virtual detector (black) relative to the bright field and diffracted disks. The virtual detector lies entirely within the bright field disk and is divided into segments. (b-c) Opposing segments are used to create virtual DPC images. (d-e) Virtual DPC images are produced using the segments defined in (b) and (c), respectively. (f-g) Line profiles of (d-e). The line profiles are comparable to Fig.~\ref{fig:LineProfSupp}.
    }
\end{figure}

To parallel our COM imaging results (Figure 2), we tested the application of differential phase contrast (DPC) imaging using a virtual DPC detector with eight segments (Figure S3a). We produced virtual DPC images from the same experimental data set as in Figure 2. For each scan position, the value in the virtual DPC image is calculated as a weighted average of the total intensity within the virtual detector segments, with the outer segments having weights of $\pm 2.28$ mrad and the inner segments $\pm 1.23$ mrad. Line profiles of these virtual DPC images (Figure \ref{fig:DPC}f-g) resemble line profiles of center of mass images prior to background subtraction (Figure~\ref{fig:LineProfSupp}), and again, fringes with a period of \SI{12}{\angstrom} are visible.  This is reasonable, as both DPC and COM imaging can be thought of as taking the difference of two selected regions of the diffraction pattern.  In principle, DPC-STEM can also detect the magnetic structure, but the detector must be oriented correctly and have the correct radius.  In practice, a pixelated detector with high dynamic range makes it easier to set up the experiment.

\section{Density functional theory calculations}
The charge and magnetization densities of antiferromagnetic tetragonal \ch{Fe2As}, are calculated using first-principles density functional theory (DFT) as implemented in the Vienna Ab Initio Simulation Package (VASP)\cite{Kresse1996,Kresse1999}. The electron-ion interaction is described by the projector augmented wave (PAW) method \cite{BlochlPE}. We use the generalized-gradient approximation (GGA) parameterized by Perdew, Burke, and Ernzerhof (PBE) to describe exchange and correlation \cite{PerdewJP}. Kohn-Sham states are expanded into a plane-wave basis up to a kinetic energy cutoff of 600 eV and we use a $15 \times 15 \times 5$ Monkhorst-Pack (MP) \cite{MonkhorstHJ} k-point grid to sample the Brillouin zone. These parameters lead to a total energy convergence of about 0.06 meV/atom. Energy self consistency is converged to within $10^{-6}$ eV. Simulations are performed for the fully relaxed lattice parameters of $a= \SI{3.624}{\angstrom}$, $b= \SI{3.624}{\angstrom}$, and $c= \SI{11.720}{\angstrom}$ described in Ref.~\citenum{kang2020polar}. All simulations are carried out using the noncollinear description of magnetism including spin-orbit coupling \cite{SteinerS_PRB}. The resulting charge and magnetization densities are subsequently computed on $36 \times 36 \times 120$ real-space grid points.

\section{Magnetic multislice simulations}
\subsection{Calculation of vector potential and magnetic fields}
We calculate the periodic magnetic vector potential $\vec{A}(\vec{r})$ for one magnetic unit cell of \ch{Fe2As}, calculating the value of a series of fields on the same real-space grid used by the DFT calculations.  First, given the magnetization density $\vec{M}(\vec{r})$ calculated from DFT, we calculate the current density $\vec{j}(\vec{r})$ induced by the magnetization density.  Next, according to magnetostatics, under the Coulomb gauge,
\begin{equation}
\vec{A}(\vec{r}) = \frac{\mu_0}{4\pi} \int d^3 r' \frac{\vec{j}(\vec{r}')}{|\vec{r} - \vec{r}'|}.
\end{equation}
We calculate this integral numerically over multiple magnetic unit cells in a cuboid-shaped region near $\vec{r}$, thereby including within the integral the nearest periodic images of the current density at each grid point.  This type of method of expanding rectangles for computing alternating sums has been shown to converge \cite{Borwein1998}.

Similarly, we also calculate magnetic fields in \ch{Fe2As}.  We calculate the distribution of fictitious magnetic charges induced by magnetization: $\rho^*(\vec{r}) = -\nabla \cdot \vec{M}(\vec{r})$. We can then find the magnetic field via
\begin{equation}
\vec{H}(\vec{r}) = \frac{1}{4\pi} \int d^3 r' \rho^*(\vec{r}') \frac{\vec{r} - \vec{r}'}{|\vec{r} - \vec{r}'|^3}.
\end{equation}


\subsection{Adding magnetism to multislice electron scattering simulations}

To model electron scattering in a sample with magnetic fields, Ref.~ \citenum{Edstrom2016Vortex} use the following two-component paraxial Pauli equation:

\begin{multline}
\pd{}{z} \begin{pmatrix} \psi_\uparrow(\vec{r}) \\ \psi_\downarrow(\vec{r}) \end{pmatrix} = \frac{im}{\hbar(\hbar k + eA_z)} \biggl\{ \frac{\hbar^2}{2m} \nabla_{xy}^2 + \frac{ie\hbar}{m}\vec{A}_{xy} \cdot \nabla_{xy} \\
- \frac{\hbar k e A_z}{m} - \frac{e}{m} \hat{S} \cdot \vec{B} + eV \biggr\}\begin{pmatrix} \psi_\uparrow(\vec{r}) \\ \psi_\downarrow(\vec{r}) \end{pmatrix}.
\end{multline}

Given the magnitude of $\vec{A}$, $\vec{B}$ calculated for \ch{Fe2As} and the orientation of spins within the sample, several of the terms in this equation can be neglected:
\begin{enumerate}
	\item $eA_z$ in the denominator, since $\hbar k \gg eA_z$.
	\item $\frac{ie\hbar}{m}\vec{A}_{xy} \cdot \nabla_{xy}$. Because the magnetic moments lie in the plane of the sample, $\int_z^{z + \Delta z} \vec{A}_{xy}\, dz = \vec{0}$ for an integral of one unit cell along $z$.
	\item $- \frac{e}{m} \hat{S} \cdot \vec{B}$, because it is $\approx 10^{-2}$ the size of $- \frac{\hbar k e A_z}{m}$, and because we use an unpolarized electron beam.  As a result, we need not consider two spin components.
\end{enumerate}

In summary, for our periodic magnetic structure with zero net magnetic moment and in-plane magnetic moments, we can use the following paraxial equation:
\begin{equation}
\pd{\psi(\vec{r})}{z} = \left[ \frac{i}{2k}\nabla^2_{xy} + i \left( \frac{me}{\hbar^2 k} V(\vec{r}) - \frac{e}{\hbar}A_z(\vec{r}) \right) \right] \psi(\vec{r}).
\end{equation}

This equation can be modeled using the conventional multislice method, with a transmission function that includes both the electrostatic potential and the axial component of the vector potential.  One limitation of multislice simulations is that we assumed a perfectly coherent electron probe.  Effects of partial coherence are expected to reduce the COM signal, smoothing out features and reducing contrast in the central CBED disk \cite{OxleyIncoherence}.  This makes it difficult to quantitatively relate the magnitude of COM shifts to absolute values of projected B-field. For future models of multislice with magnetism, the addition of partial coherence would give us a more accurate description of the interactions of the electron beam and sample. 

\subsection{4D-STEM simulations}
The 4D-STEM simulations shown in Figure 4 and Figure 5a-g were performed with an accelerating voltage of 120 kV and semi-convergence angle of 3.4 mrad, in order to match experimental conditions. The frozen phonon approximation was used: each diffraction pattern was calculated by incoherently averaging the diffraction patterns produced from 10 phonon configurations.  In each phonon configuration, the Fe and As atoms were randomly displaced according to normal distributions with standard deviations of $\SI{5.4}{pm}$ and $\SI{3.6}{pm}$, respectively, consistent with neutron scattering results from Ref. \citenum{Rancourt1985}. In addition to the magnetic and non-magnetic Magnstem simulations shown in Figure 4, a third simulation was performed where the magnetic moments were reversed -- this was accomplished by negating the magnetic phase shift used in the simulation.

\subsection{Principal component analysis of simulated 4D-STEM datasets}

\begin{figure}
    \includegraphics[width=.85\textwidth]{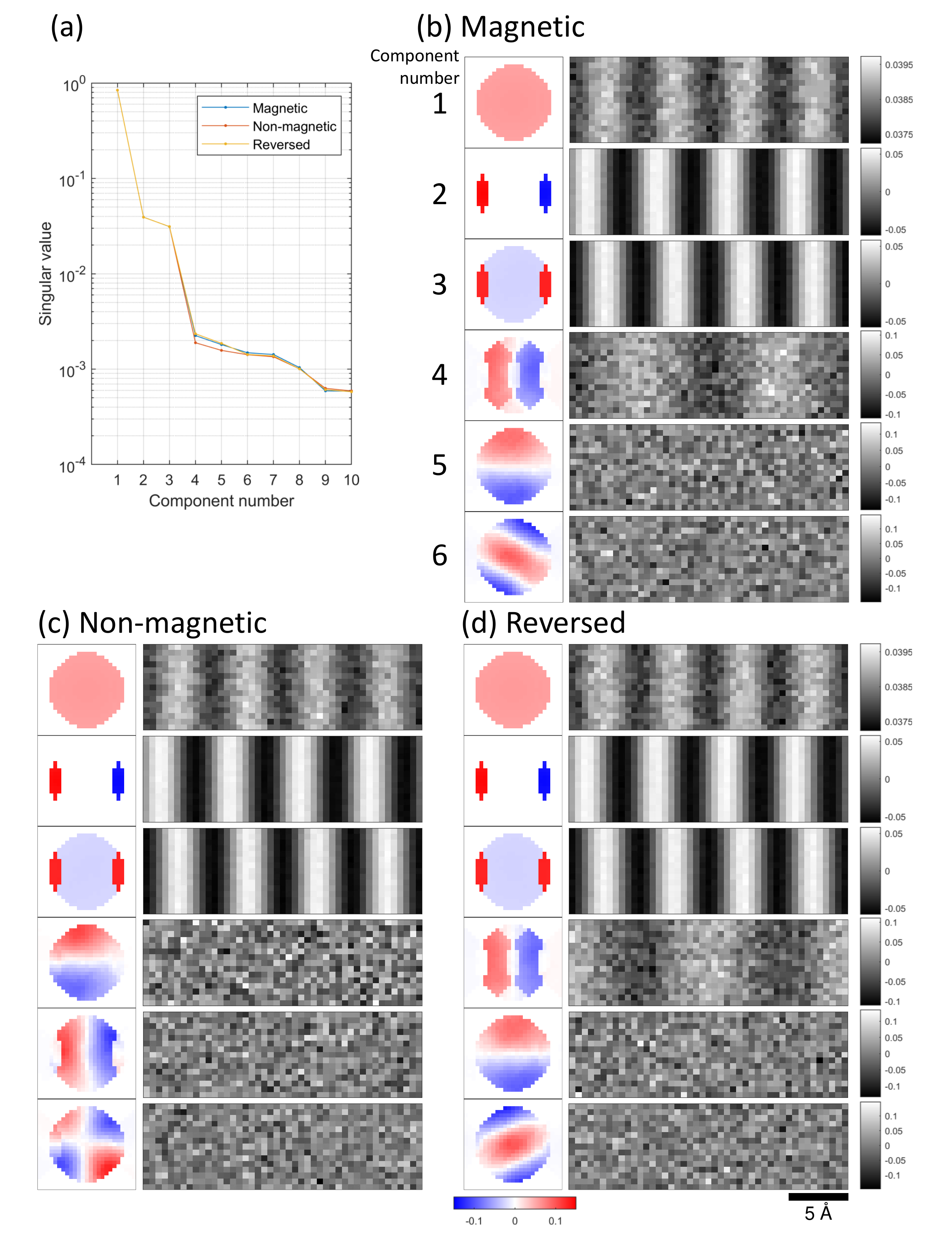}
    \caption{\label{fig:pca_separately}Principal component analyses separately on three simulated 4D-STEM data sets. (a) Scree plot of PCA components. (b) Components of the CBED pattern and corresponding real space maps, shown for the six components with largest singular values from the three simulated 4D-STEM data sets with magnetism respectively turned on, turned off, or reversed. The real space maps show the right singular vectors of the principal component analysis, rearranged to correspond to probe position. All components of the CBED pattern are plotted on the same color scale.
    }
\end{figure}

To prepare a simulated 4D-STEM data set for PCA, each diffraction pattern was unrolled into a vector, and each vector put in order by scan position, to form a matrix.  PCA was performed via singular value decomposition of this matrix. For each principal component of the CBED pattern (the load in our PCA analysis), there is a corresponding real space image corresponding to the PCA score (Figure \ref{fig:pca_separately}).  Starting from the fifth principal component onwards, all principal components real space maps that showed only noise, arising from random atomic displacements from the frozen phonon approximation.

\subsection{Size of magnetic deflections as a function of thickness}
To quantify the effects of sample thickness, we conduct multislice simulations of \ch{Fe2As} up to a sample thickness of 150 unit cells (54.3 nm), scanning the probe over 8 scan positions along the center of the three-row block. In those scan positions, the probe is mostly averaging over a region of aligned spins within a block, thus maximizing the effect of the magnetic structure on the CBED pattern. We use a semi-convergence angle of 3.4 mrad.  CBED patterns are stored at thicknesses from 1 unit cell (0.36 nm) to 150 unit cells (54.3 nm).  CBED patterns are averaged over 90 phonon configurations.

\subsubsection{Definition of ``projected B field'' in Figure 5h}
As described in Ref.~\citenum{Rusz_QuantumDPC}, in DPC and COM measurements, the mechanical momentum displacement vector $\vec{P}$ is approximately evaluated as
\begin{equation}
    \langle \vec{P}_\perp \rangle = \sum_n \vec{P}_{\perp, n} w_n.
\end{equation}
Each segment or pixel of the detector is associated with some vector $\vec{P}_{\perp, n}$, and $w_n$ is the intensity collected by that segment or pixel.  Electric and magnetic fields in the sample cause shifts in this signal $\Delta \langle \vec{P}_\perp \rangle$.  Importantly, the measured $\Delta \langle \vec{P}_\perp \rangle$ at some probe position $(x,y) = \vec{R}$ depends on the values of the fields not just at $\vec{R}$, but over an extended region centered on $\vec{R}$, according to the distribution of the electron probe's intensity.

However, to obtain a simple estimate of how much the electron beam interacts with the magnetic fields of a sample, we ignore how the intensity distribution of the electron probe is complicated by diffraction and beam spreading, and define the projected B-field as 
\begin{equation}
    \tilde{\vec{B}} = \frac{1}{A_\perp} \int_V \vec{B}(\vec{r})\,d^3\vec{r}.
\end{equation}

Here, $V$ is a cylindrical volume whose axis is the path of the beam within the sample, $A_\perp$ is the cross-sectional area of $V$, and we use the magnetic field $\vec{B}(\vec{r})$ calculated from density functional theory for one magnetic unit cell.  In our STEM experiments, the grain of \ch{Fe2As} is oriented such that the $c$ axis points in the $x$-direction.  The electron beam travels along the $z$-direction, causing the magnetic moments of the sample to point along the $y$-direction. If the electron beam lies within one trilayer of \ch{Fe2As}, it will interact with magnetic moments all pointing in the same direction, leading to a greater magnetic signal. We assume $V$ to be a cylinder with radius \SI{3}{\angstrom} and height equal to the thickness of the sample. When $V$ is centered on a trilayer of \ch{Fe2As} and extends through a sample with a thickness of $t=$ 1 unit cell (\SI{3.62}{\angstrom}), $\tilde{\vec{B}}$ points in the $y$ direction and has a magnitude of $\tilde{B}_{\rm{1uc}} = \SI{3.9}{\tesla.\angstrom}$.  Assuming a periodic magnetic structure, $\tilde{\vec{B}}$ grows linearly with sample thickness.  Hence we plot lines in Figure 5h with slope equal to $\pm\tilde{B}_{\rm{1uc}}/t$.

\subsection{Effects of semi-convergence angle}

\subsubsection{Multislice simulations}
To understand the effects of semi-convergence angle, we performed a series of multislice simulations of \ch{Fe2As} with semi-convergence angles ranging from 1.6 mrad to 4.6 mrad.  For simulations with semi-convergence angle $\leq 2.8$ mrad, a supercell of size \SI{203.0}{\angstrom} $\times$ \SI{187.5}{\angstrom} and a probe wavefunction with $4096 \times 4096$ pixels was used; for simulations with semi-convergence angle $\geq 3.0$ mrad, a supercell of size \SI{101.5}{\angstrom} $\times$ \SI{93.8}{\angstrom} and a probe with $2048 \times 2048$ pixels was used.  The simulated sample had a thickness of 9.1 nm (25 unit cells).  The probe was rastered over 15 x 45 positions over the sample. At each semi-convergence angle, three simulations were performed: with magnetism turned on, turned off, and reversed in direction.  Resulting CBED patterns were cropped to squares with a width that is 20\% greater than the diameter of the BF disk.  For each set of three simulations at a particular value of semi-convergence angle, we performed principal component analysis on the cropped CBED patterns.  Unlike the PCA analysis shown in Figure 4 and Figure S4, the PCA analysis here is performed only once for the set of three simulations -- CBED patterns from all three simulations are included in the matrix prior to SVD.

In order to reduce computation time, the frozen phonon model was not used; atoms were not randomly displaced. Although the lack of thermal diffuse scattering in these simulations does slightly affect the intensity of the Bragg disks, the simulated sample is thin enough that the effect of thermal diffuse scattering is small compared with the interference of the magnetic \zerohalf disks with the BF disk.  Thus, we do not expect not using the frozen phonon model to affect our conclusions.

\begin{figure}
    \includegraphics[width=.5\textwidth]{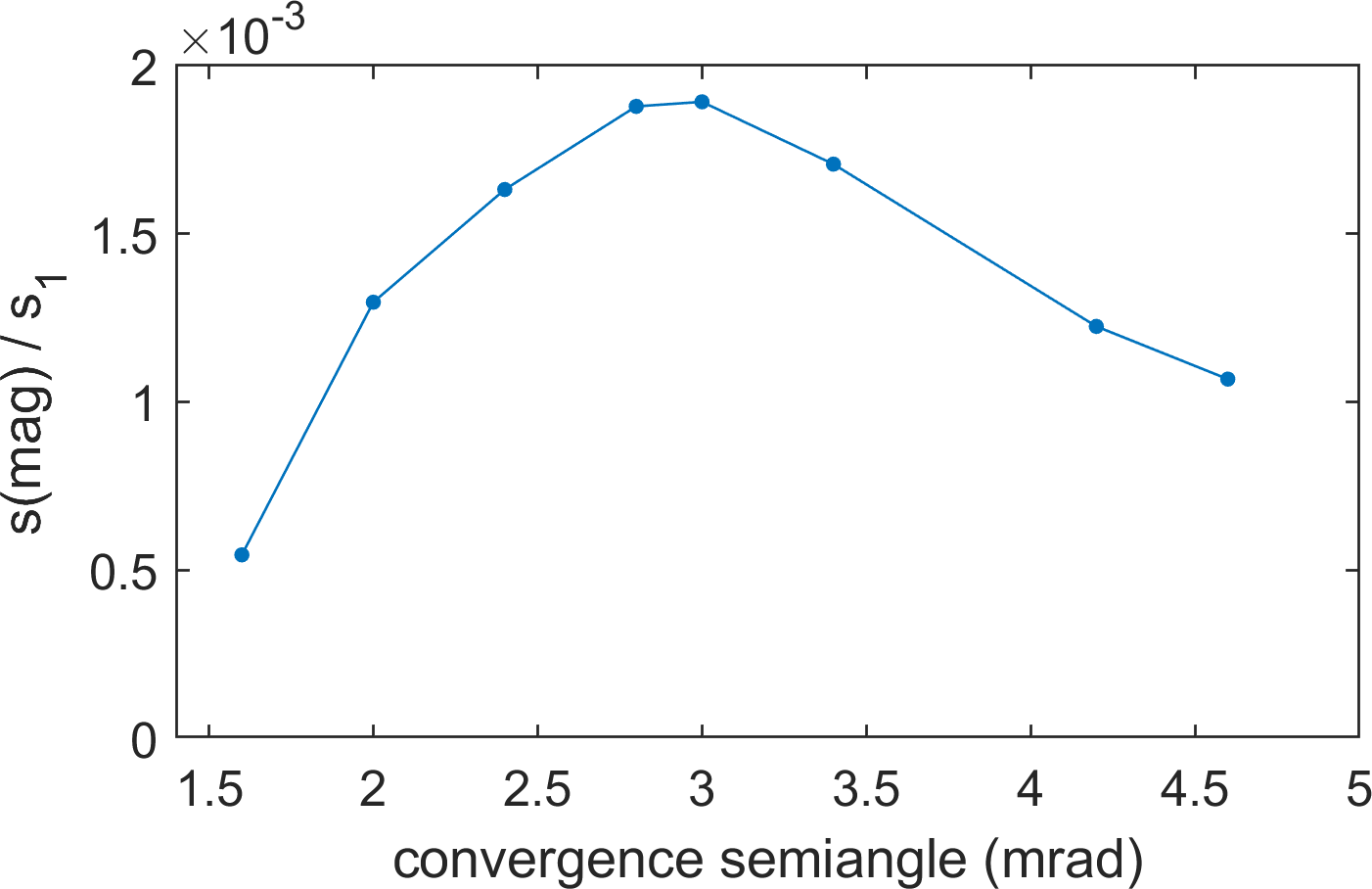}
    \caption{Ratio of singular values of magnetic PCA component and largest PCA component}
    \label{fig:s_ratios}
\end{figure}

For each semi-convergence angle value, the largest principal component of the CBED pattern is the average CBED pattern, while one of the smaller principal components shows the overlap of \zerohalf disks with the BF disk, corresponding to the magnetic structure.  We compute the ratio of singular values of the magnetic PCA component to that of the largest PCA component; this is shown in Figure~\ref{fig:s_ratios}. The ratio peaks at 3.0 mrad with a value of $1.9 \times 10^{-3}$.  This gives an estimate of the size of the effect that the magnetic structure of \ch{Fe2As} has on 4D-STEM data sets.

\subsubsection{Optimizing the overlap of Bragg disks}

\begin{figure}
    \includegraphics[width=.48\textwidth]{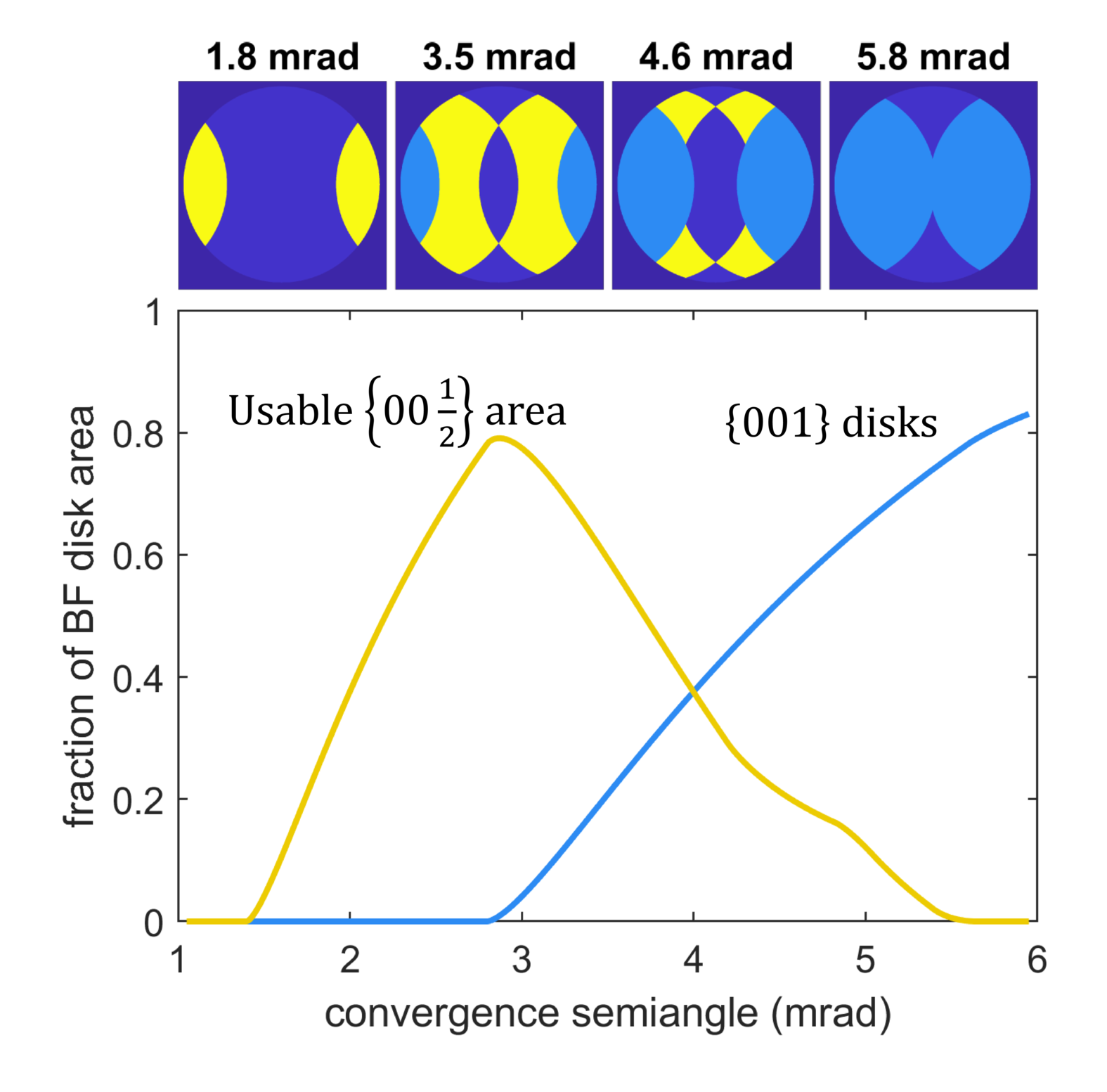}
    \caption{Effect of convergence semiangle on visualizing \zerohalf lattice fringes. (top) Schematics of the BF disk drawn for four different values of convergence semiangle.  The yellow regions show the ``usable'' \zerohalf area, defined as the regions where the \zerohalf disks overlap with the BF disk but not with any other Bragg reflections of \ch{Fe2As}. The $\{ 001 \}$ disks are shown in light blue. (bottom) The size of the usable \zerohalf area and $\{ 001 \}$ disks, as a fraction of the BF disk, are plotted as a function of the convergence semiangle.}
    \label{fig:conv_angle_areas}
\end{figure}

For an accelerating voltage of 120 kV, the semi-convergence angle must be greater than 1.4 mrad for the \zerohalf disks to overlap with the BF disk, and less than 5.6 mrad for the \zerohalf disks not to be completely covered by other structural Bragg disks (see Figure~\ref{fig:conv_angle_areas}). At higher semi-convergence angles where the $\{ 001 \}$ disks completely overlap with the \zerohalf antiferromagnetic disks, the magnetic structure still produces slight intensity changes in the BF disk, but those changes are more challenging to measure and isolate, since the disks and dynamical scattering from the atomic structure make the contrast within the BF disk more complicated.

 \bibliographystyle{elsarticle-num} 
 \bibliography{fe2aspaper}




